\documentclass[apl,superscriptaddress,reprint,amssymb,aps,floatfix,showkeys]{revtex4-1}
\usepackage{amsmath}%
\usepackage{amsfonts}%
\usepackage{amssymb}%
\usepackage{graphicx}
\usepackage{color}
\usepackage{tabularx}
\usepackage{braket}

\begin{document}


\title{Ultraviolet Light-Induced Microwave Mode Tuning in a Rutile TiO$_2$ Whispering Gallery Resonator}
\author{Catriona A. Thomson}
\email{catriona.thomson@research.uwa.edu.au}
\affiliation{Quantum Technologies and Dark Matter Labs, Department of Physics, University of Western Australia, 35 Stirling Highway, Crawley, WA 6009, Australia.}
\affiliation{Humboldt Centre for Nano- and Biophotonics, Institute for Light and Matter, Department of Chemistry and Biochemistry, University of Cologne, Köln, Germany.}
\author{Michael E. Tobar}
\affiliation{Quantum Technologies and Dark Matter Labs, Department of Physics, University of Western Australia, 35 Stirling Highway, Crawley, WA 6009, Australia.}
\author{Maxim Goryachev}
\email{maxim.goryachev@uwa.edu.au}
\affiliation{Quantum Technologies and Dark Matter Labs, Department of Physics, University of Western Australia, 35 Stirling Highway, Crawley, WA 6009, Australia.}
\date{\today}

\begin{abstract}
We report the observation of transient nonlinear optical effects in a macroscopic whispering gallery mode resonator made of rutile TiO$_2$, demonstrating strong optical-microwave transduction under laser irradiation. By comparing the effects of ultraviolet (UV, 385 nm) and near-infrared (NIR, 700 nm) radiation, we find that the UV-induced effects are significantly amplified, consistent with the material’s semiconductor bandgap energy. The interaction results in frequency shifts of microwave modes and changes in quality factor, suggesting a localized saturable refractive index tuning. This may be attributed to the saturation of a spin transition of a dopant ion within the crystal lattice. Remarkably, these effects are observed at low optical powers, down to nanowatt levels, indicating high sensitivity and efficient of light-matter interaction in this system. The phenomenon is centered around 15 GHz, yet electron spin resonance measurements reveal no zero-field splitting at this frequency, suggesting an alternative mechanism beyond conventional spin resonance. These findings highlight the potential of low-power optical control of microwave modes in high-Q resonators for applications in quantum technologies, sensing, and reconfigurable photonic-microwave systems.
\end{abstract}

\maketitle

\section{Introduction}

The field of microwave photonics, the transduction of optical and microwave signals, has been flourishing in recent years due to its high relevance to high-bandwidth communications and various sensing applications, and is bourgeoned by abundant research in nonlinear optical materials~\cite{Marpaung2019,Lin2017,Ilchenko2006}. Photosensitive semiconductor and dielectric materials are particularly useful for such applications, owing to intrinsic nonlinear effects such as Pockels, Kerr, Raman and Brillouin~\cite{Lin2017}, which may be harnessed to enable momentum and energy transfer between light waves at different frequencies~\cite{Ilchenko2006}, create frequency combs~\cite{Wu2018}, and enable all-optical switching~\cite{Yuece2016}. Significant effects upon the electromagnetic resonance properties of solid state cavities which host paramagnetic impurities have also been found when the energy level occupancy in these dopant impurities is manipulated~\cite{Farr2013}.

Whispering-gallery mode (WGM) resonators offer an ideal platform for both material characterization and applied microwave photonics at very low threshold powers~\cite{Sun2017,Li2020} due to their high quality factors which locally confine light for exceptionally long periods, enabling significant power build-up, greatly enhancing light-matter interactions~\cite{Jiang2020B,Lin2017}. WGMRs were first used in fundamental studies of nonlinear optics and laser physics, but their valuable enhancement of intrinsic nonlinear effects and sensitivity to external refractive index changes has seen their applications rapidly diversify in the last 20 years, from biomedical applications in in-vitro and in-vivo biosensing~\cite{Toropov2021,Schubert2015,Schubert2020,Titze2022}, to electric and magnetic field sensing, gas, pressure and temperature sensing~\cite{Jiang2020B, Park2014}, microwave switching, studies of quantum electrodynamics~\cite{Kosma2014,Jiang2020,Lin2017}, as well as tunable filtering and laser stabilisation~\cite{Ilchenko2006}. Significantly, harnessing their bulk linearities, WGMRs have the potential to act as all optical components in telecommunications systems, avoiding the bandwidth limitation inherent in optoelectronic information transfer~\cite{Lin2017,Jiang2020}.

Apart from an abundance of studies on nonlinearities in one-dimensional microcavities and waveguide devices~\cite{Yuece2016,Hagn2001,Trocha2021}, a great amount of research has been made into nonlinear behaviour within WGM resonators of various structures including toroids, spheres and rods~\cite{Heylman2017}, in particular based upon silica (SiO$_2$), calcium fluoride (CaF$_2$), lithium niobate (LiNbO$_3$)~\cite{Matsko2006,Wu2018,Lin2017}, sapphire (Al$_2$O$_3$)~\cite{Ilchenko2014,Thomson2022}, gallium arsenide (GaAs), gallium phosphide (GaP)~\cite{Krupka2008} and gallium nitride (GaN)~\cite{Li2018}.

Titanium dioxide (TiO$_2$) is a wide band gap semiconductor ripe for exploitation as a microwave photonic transducer. The rutile phase of TiO$_2$ has a relatively high refractive index (2.7), compared to silicon dioxide (1.45) and a large bandgap ($E_g =$ 3.1 eV), with low material loss in the visible to infrared band, 10 times lower than that of silicon at 1.5 $\mu m$ \cite{Wang2005}. Interestingly, TiO$_2$ has a low thermal expansion coefficient, but a significant negative thermo-optic coefficient, which is fairly rare among CMOS compatible photonic materials \cite{Park2014}. TiO$_2$ also displays a host of nonlinear phenomenon, including the highest known birefringence of all naturally occurring dielectric crystals~\cite{Ha2019}, and a strong Kerr nonlinearity ($n_2 = 9 \times 10^{-19}$ m$^2$/W)~\cite{Evans2012}, 30 times larger than silica which motivates its use in nonlinear optical devices. Its transparency at the telecom wavelengths ($>$400 nm) motivates its use as a material for all-optical switching and logic applications~\cite{Evans2012}.

Due to its high photocatalytic activity, biocompatibility and memristic capability, there has been significant research into TiO$_2$ nanoparticles and thin films for industrial sterilization, UV protection~\cite{Haider2019}, photodynamic therapy in medicine~\cite{Ziental2020} and resistive switching applications~\cite{Illarionov2020}. However, its study in the context of resonantly enhanced microwave photonics, in particular in the whispering gallery mode configuration, is relatively underdeveloped.

Rutile, which is the centrosymmetric tetragonal form of TiO$_2$, and the form used in this work, has previously been studied for its nonlinear phenomenon, including third harmonic generation~\cite{Gravier2007,Borne2012}, four-wave mixing~\cite{Vianna1986} and hyperparametric oscillation~\cite{Nand2014}. In~\cite{Park2014}, Park et al. observed optically-induced wavelength tuning and broadening of optical WGMs in novel rutile micro-goblet resonators due to nonlinear phenomenon.

In this paper, we report the first observations of optically-induced tuning of WGM microwave resonances in rutile, across a significant portion of frequency space (5 - 22 GHz), displaying microwave-frequency and optical-frequency dependant refractive index tuning within the crystal resonator. The magnitude of frequency shift induced by constant radiation at either 700 nm or 385 nm was found to nonlinearly depend upon the intensity of applied laser light. The direction and magnitude of shift was found to be highly dependent upon the microwave mode of interest, showing a crossover point in shift direction at about $15$ GHz. 

\section{Experimental Setup}

The experiment is built around a single crystal rutile cylinder ($\langle001\rangle$ orientation) of $7$ mm diameter and $3$ mm thickness with a central punch hole of $1$ mm diameter.  The crystal is housed in a fully enclosed metallic (non-superconducting) cavity and sits on a metallic post support going through the central hole. Microwave modes inside the crystal are coupled to using standard microwave loop probes sitting underneath the crystal. It is irradiated with NIR/UV light injected into the cavity from side windows via the exposed end of a standard multimode fibre. 

\begin{figure}[t!]
\center
\includegraphics[width=0.71\columnwidth]{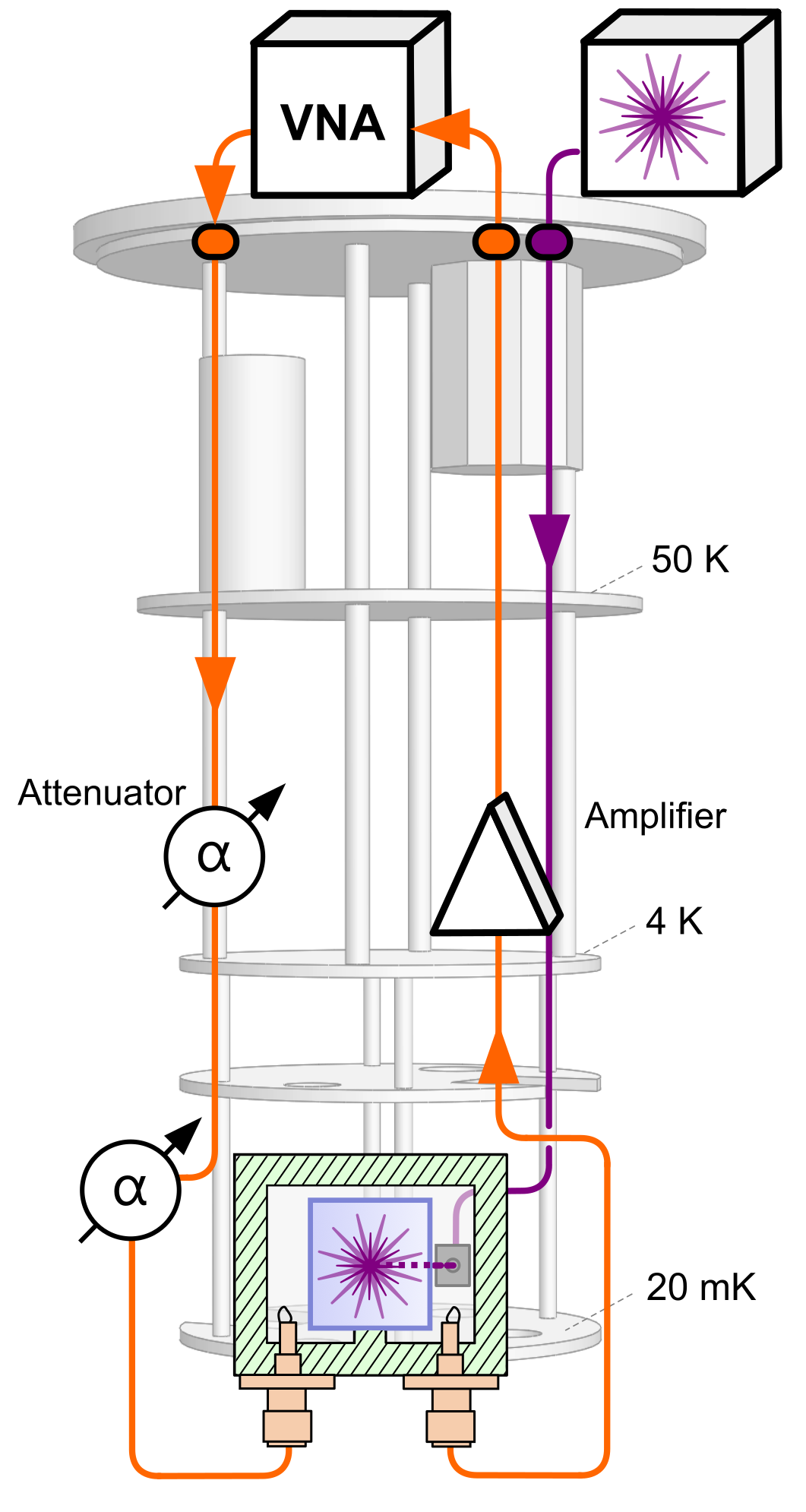}
\caption{Experimental setup showing a rutile WGM cavity sitting at the 20 mK stage of a dilution refrigerator. The system is characterised by a microwave circuit consisting of a chain of cold attenuators and a cold amplifier. The NIR/UV light is injected via fibre from an LED.}
\label{setup}
\end{figure}

The assembly sits on a $20$ mK plate of a dilution refrigerator coupled to a microwave chain consisting of s series of cold attenuators ($-10$ dB at 4 K and $-20$ dB at 20 mK) and a cold amplifier sitting at 4 K. The resonator is characterised with a vector network analyser locked to a stable frequency reference source. The light is injected into the cavity via an optical fibre sourced from a room temperature controlled LED. As a UV source, the LED's central wavelength is chosen to be 385 nm. As a test signal another LED with central wavelength of 700 nm is used. The experimental setup is shown in Fig.~\ref{setup}.

The Electron Spin Resonance (ESR) spectroscopy of the sample is performed using a 7 T superconducting magnet. The spectroscopy was based on the same cavity cooled to 4 K using the same microwave measurement chain. The external field $B$ was applied along the resonator's longitudinal axis. This technique has been used previously to study paramagnetic impurities in low loss crystals at low temperatures~\cite{Farr2013}. 

\section{Sensitivity Results}

\begin{figure}[!t]
\center
\includegraphics[width=0.8\columnwidth]{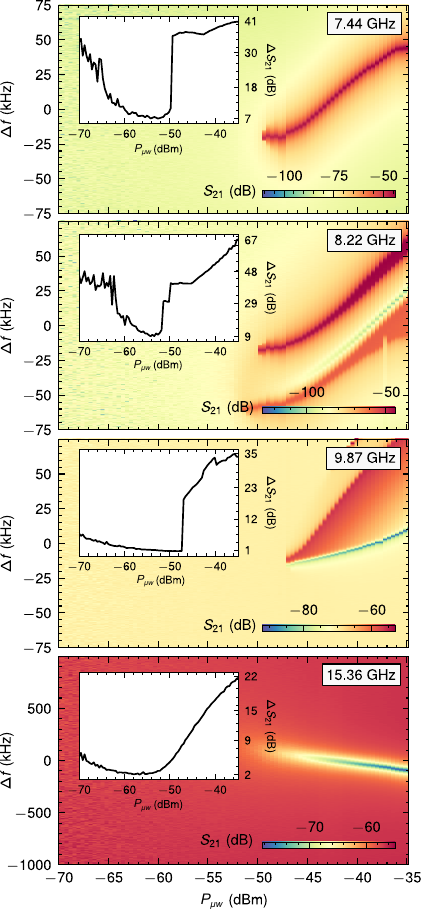}
\caption{Microwave power sensitivity of selected WGMs in a rutile crystal at 20 mK as shown by the microwave transmission coefficient $S_{21}$ as a function of incident microwave power $P_{\mu W}$. Inserts show mode contrast $\Delta S_{21}$ for given power as given by Eq. (\ref{contrast}), showing a threshold for all modes near -50 dBm.}
\label{microwavedep}
\end{figure}

Previous tests of rutile based microwave WGM resonators at cryogenic temperatures revealed significant nonlinear effects~\cite{Nand2014}. In this study, we confirm strong incident power dependence of high-$Q$ modes in the crystal. This is done by measuring crystal microwave transmission as a function of incident power $P_{\mu W}$. The results are shown in Fig.~\ref{microwavedep} for four modes. For all cases, the crystal modes do not exist for low excitation power, typically below a threshold power of around $-50$ dBm. This is clearly detailed in the inserts, which plot mode contrast defined as the difference between maximum and minimum values of the transmission coefficient for given power, $P$:
\begin{equation}
\Delta S_{21}(P) = \max_{f} |S_{21}(f,P)| -  \min_{f}  |S_{21}(f,P)|.
 \label{contrast}
\end{equation}
Typically this type of nonlinearity is due to saturable absorption when a finite ensemble of two level systems (TLS) is coupled to photonic modes. Thus, for low stored energy, the ensemble works as an additional electromagnetic loss mechanism. For larger powers, the TLS ensemble saturates and the corresponding loss becomes negligible, dominated by other mechanisms. For the microwave frequency range, this TLS induced nonlinear loss mechanism becomes prominent at temperatures below 1 K where the thermal population of the TLS ensemble transitions to the ground state. This effect has been observed in Al$_2$O$_3$ WGM resonators containing Fe$^{3+}$ or Cr$^{3+}$ ions pumped at higher ion transitions~\cite{Creedon:2011aa}.

\begin{figure}[!b]
\includegraphics[width=0.8\columnwidth]{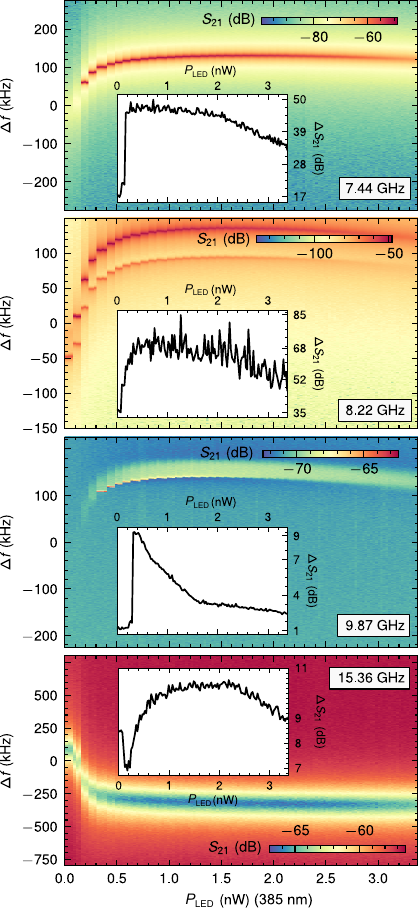}
\centering
\caption{Dependence of the WGM resonator transmission on power of incident UV light as shown by microwave transmission coefficient $S_{21}$ as a function of incident LED power $P_\text{LED}$. Inserts show mode contrast $\Delta S_{21}$ for given power.}
\label{uvdep}
\end{figure}

\begin{figure*}[!t]
\center
\includegraphics[width=1.9\columnwidth]{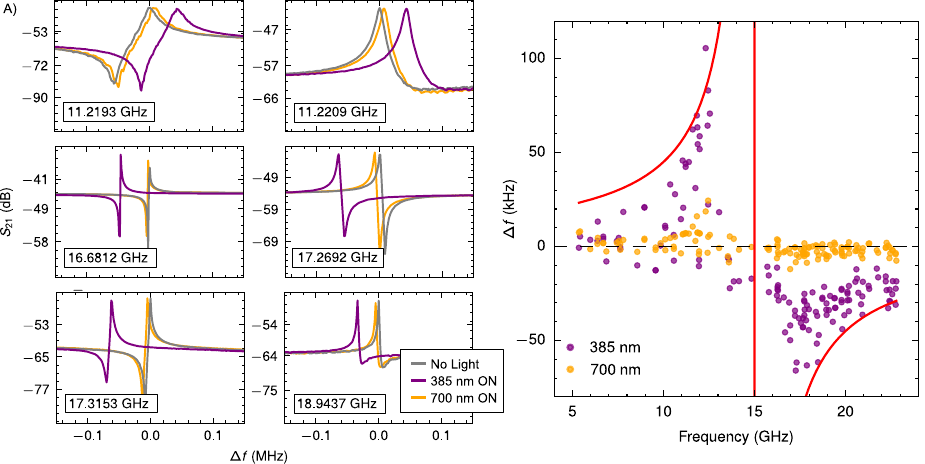}
\caption{(A) Examples of change of transmission spectra for 6 modes. (B) Frequency shifts $\Delta f$ of WGMs frequencies as a function of resonance frequency $f$ for 385 nm and 700 nm illumination. The red line indicates the frequency shifts expected due to the susceptibility added by an unidentified paramagnetic ion with a transition frequency at $15$ GHz, a dc susceptibility of $\chi_{\gamma} = 2 \times 10^{-6}$ and a spin-spin relaxation time of $\tau_0 = 5 \times 10^{-8}$, assuming unity magnetic filling factor.}
\label{fvdf}
\end{figure*}

To demonstrate the sensitivity of rutile to incident UV light under cryogenic conditions, a set of transmission spectra were measured as a function of LED power $P_\text{LED}$. The result is shown in Fig.~\ref{uvdep} where cases below and above the nonlinear threshold are depicted. In all cases, one observes both frequency shifts $\Delta f$ (up to hundreds of kHz) and transmission enhancement as shown in inserts using mode contrast $\Delta S_{21}$. For the case below the observation threshold, modes appear around $0.25$ nW. This result suggests that UV light alters both real and imaginary part of the effective permeability, the former leads to a shift in resonance frequency, and the latter in an increase of effective $Q$ value and as a result in increased peak transmission. Note that $\Delta S_{21}$ is directly proportional to the loaded $Q$ factor of a resonance.

At larger UV power ($2$ nW and above), one observes a decrease in transmission amplitude and opposite frequency shifts. This might be explained by additional heating effects from the incident light. Overall, this effect also suggests the existence of saturable loss channel, most likely an ensemble of TLS.

A frequency shift $\Delta f$ between dark and illuminated crystal is measured for all high $Q$ WGMs in the frequency range from 5 to 24 GHz. In this case, the highest incident optical power is used. Fig.~\ref{fvdf} (A) gives examples of transmission for 6 modes for unilluminated, 385 nm and 700 nm illuminated cases. Comparing the two wavelengths, it is clear that only the UV illumination gives substantial frequency shifts excluding generic optical heating or photothermal effects.

Indeed, the absence of an effect at 700 nm (~1.77 eV) suggests that the observed frequency shift and $Q$-factor increase are not due to general optical heating but require higher-energy photons. This points to a mechanism involving bandgap transitions or defect activation. Since rutile TiO$_2$ has an indirect bandgap of $\sim 3.0$eV ($413$ nm) and a direct bandgap of $\sim 3.3$ eV ($375$ nm), 700 nm light is insufficient to excite electrons across the bandgap. However, UV light ($385$ nm or shorter) provides enough energy to generate charge carriers, potentially altering the dielectric properties of the material.

Frequency shifts $\Delta f$ over the whole frequency range for the two wavelengths are summarised in Fig.~\ref{fvdf} (B). The result shows that both positive and negative $\Delta f$ are observed for the UV case with the clear border between them at around 15 GHz which also excludes broadband heating or optical effects. The two signs can be attributed to negative and positive signs of the effective permeability change. Similar effects have been observed in Al$_2$O$_3$ containing Fe$^{3+}$ and Cr$^{3+}$ ions~\cite{Tobar2003,Benmessai:2009aa}. In the case of Fe$^{3+}$, the shifts are controlled via pumping of higher level spin transitions. 

The complex microwave susceptibility of a solid altered by a background TLS resonance in the crystal lattice can be expressed by:
\begin{equation}
\chi^{\prime}(\omega)+j \chi^{\prime \prime}(\omega) = \frac{\tau_0^2 \omega_0 \chi_{\gamma}\left(\omega-\omega_0\right)}{1+\tau_0^2\left(\omega-\omega_0\right)^2}-j \frac{\tau_0 \omega_0 \chi_{\gamma}}{1+\tau_0^2\left(\omega-\omega_0\right)^2}
\label{susceptibility}
\end{equation}
where $\omega = 2\pi f$ is the probed frequency,  $\chi_{\gamma}$ is the static magnetic susceptibility, $\omega_0 = 2\pi f_0$ is the angular frequency of the spin transition and $\tau_0$ is the relaxation time of the transition~\cite{Mann2000}. 

The static magnetic susceptibility may exhibit anisotropy with respect to the crystal axis which we do not consider in this work. This susceptibility change results in a frequency shift that can be related to the real part of the added susceptibility via
\begin{equation}
\Delta f= \frac{p_{m \gamma}}{2} \chi^{\prime}(f) f_0
\label{shiftrelation}
\end{equation}
where $f$ is the probed frequency, $f_0$ is the frequency of the TLS transition, $\Delta f = \frac{f-f_0}{f_0}$, and $ p_{m \gamma}$ is the mode-dependant magnetic filling factor (between 0 and 1) parellel to the axis in which the paramagnetic ion is oriented~\cite{Creedon:2011aa}. As measured whispering gallery modes are not necessarily oriented parallel to the spin axis, $ p_{m \gamma}$ may be reduced for some cases. This model has been successfully used to describe magnetic impurities in crystals. We apply this model to described observed frequency shifts due to application of the UV light.

Fig.~\ref{fvdf} (B) displays the experimental frequency shifts observed upon illumination of the sample by 385 nm light compared with the theoretical frequency shifts which would be expected to occur due to added susceptibility imparted by a saturated TLS transition with a frequency of $15$ GHz, a DC susceptibility of $\chi_{\gamma} = 2 \times 10^{-6}$ and a spin-spin relaxation time of $\tau_0 = 5 \times 10^{-8}$, assuming unity magnetic filling factor. A thorough examination of the literature could not identify a zero-field transition corresponding to this frequency among common doping ions of rutile titanium dioxide such as chromium, iron or manganese~\cite{Lampe:1966aa,Guler:2010aa}.

\begin{figure}[!t]
\includegraphics[width=0.95\columnwidth]{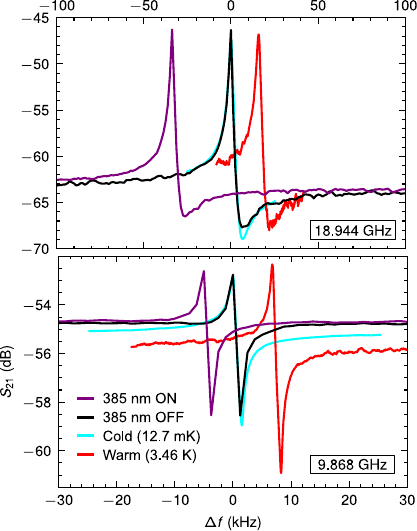}
\centering
\caption{Comparison of resonance profiles at different temperatures (12.7 mK and 3.46 K) and illumination conditions (no light and 385 nm light applied at a power of 1.24 $\mu$W).}
\label{tdepen}
\end{figure}

In order to further confirm that the observed frequency shifts are not simply due to light induced temperature variations, one can estimate frequency-temperature sensitivity based on measurements at $12.7$~mK and $3.46$~K with and without illumination. Transmission spectra for these four cases are shown in Fig.~\ref{tdepen} for two modes. Examining the 19 GHz mode, the sensitivity of the mode frequency to temperature is $\frac{\Delta f}{\Delta T}\sim 4.58$ kHz/K. One can also examine the frequency shift upon applied laser light at 385 nm at a power of 1.24 $\mu$W. The observed optical power sensitivity is $\frac{\Delta f}{\Delta P_\text{LED}}\sim -27.2$ kHz/$\mu$W. As the mode shift is saturated at an applied power far lower than 1.24 $\mu$W, it is inaccurate to derive this linear relation. However, if one assumes that the frequency shift due to applied light is temperature-induced, it would be necessary for $\frac{\Delta P_\text{LED}}{\Delta T} \sim -10$K$/\mu$W to be negative, which is impossible as it implies cooling with light. Therefore, we deduce that the frequency shift due to the applied light is not a heating effect and likely instead induced by impurity in the crystal lattice.

\section{ESR Spectoscopy}

To investigate whether the observed susceptibility change is due to paramagnetic impurities or other types of TLS, a series of Electron Spin Resonance spectroscopy measurements were done employing WGMs as sensitive probes of spin-photon coupling. This technique relies on measurements of microwave transmission coefficients near WGM resonance as a function of externally applied DC magnetic field $B$ and identification of resonance frequency deviations or changes in transmission amplitude. Each significant deviation of frequency and/or amplitude could be attributed to an avoided level crossing, a single on-resonance coupling between impurities and photons.

\begin{figure}[]
\center
\includegraphics[width=1\columnwidth]{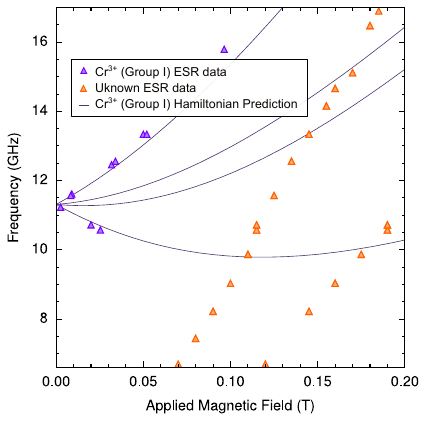}
\caption{Results of the ESR spectroscopy of the rutile crystal at 4 K together with theoretical prediction for Cr$^{3+}$ group I~\cite{Ikebe:1969aa}. }
\label{ESR}
\end{figure}

This type of ESR multimode spectroscopy has been successfully used to study impurities in various high quality crystals such as Al$_2$O$_3$~\cite{Farr2013}, quartz~\cite{Goryachev:2013aa}, YSO~\cite{Goryachev:2015aa}, YVO$_4$~\cite{Zhao:2017aa}, etc. Spectroscopy results for the rutile crystal near the frequency of maximum light sensitivity are shown in Fig.~\ref{ESR} where each point shows an identified variation in resonance parameters. The result suggests the existence of a spin system with a
zero field splitting (ZFS) near $f_\text{ZFS}= 11.31$ GHz.

According to the work of Ikebe et al, Cr$^{3+}$ spins in TiO$_2$ can be categorized into three groups; regular substitutional Cr$^{3+}$ on the Ti$^4$ sites and Cr$^{3+}$ groups I and II, representing ions substituting Ti$^4$ but in a distorted crystalline lattice caused by coupling to vacancies in the second and third oxygen sites respectively~\cite{Ikebe:1969aa}. The three groups were analysed assuming a spin Hamiltonian of S = 3/2 for an orthorhombic crystalline field. According to the group orientation and spin Hamiltonian parameters discovered in their work, we have identified one spin resonance in our data as group I Cr$^{3+}$. For this ion substitution, the y-axis coincides with the $\langle001\rangle$ direction, and the z- and x- axes are rotated from the $\langle110\rangle$ and $\langle\overline{1}10\rangle$ axes respectively by 3$^\circ$ in the (001) plane. The spin Hamiltonian parameters are $g = 1.98$, $D = 5.1$ GHz and $E = -1.41$ GHz. To determine the energy level transitions associated with this ion group, we used the EasySpin package for MATLAB~\cite{stoll2006easyspin} to perform Hamiltonian diagonalization using the aforementioned parameters and Euler angles of [90$^\circ$, 90$^\circ$, 3$^\circ$] to describe the molecular frame's inclination to the magnetic field, while assuming the magnetic field is axial to the crystal. The calculated ESR lines are shown in Fig.~\ref{ESR}, along with the spectroscopy data. The experimental data strongly suggests the presence of this Cr$^{3+}$ group I ion, with an especially good fit near zero-field. Another distinct ion group was revealed via ESR spectroscopy (Fig.~\ref{ESR}, orange data); however, this group could not be convincingly identified as any dopant of rutile TiO$_2$ previously published in the literature.

Overall these results reveal no paramagnetic impurity centred around $15$ GHz as would be expected from results in Fig.~\ref{ESR} (B) measured at zero magnetic field. 

\section{Conclusion}

This work demonstrates the control of microwave WGMs in crystalline rutile TiO$_2$ at cryogenic temperatures using low-intensity UV light. We show that 385 nm incident light can shift resonance frequencies by more than 100 kHz (hundreds of linewidths) and enhance quality factors. However, at higher optical pumping intensities, both the frequency shift and $Q$-factor improvement degrade. This behavior suggests a mechanism involving TLS with a characteristic frequency around 15 GHz.
The lack of an effect at 700 nm ($\sim1.77$ eV) indicates that the observed frequency shift and Q-factor increase are not due to general optical heating but instead require higher-energy photons, pointing to a process related to bandgap transitions or defect activation. UV light may depopulate TLS states or alter their energy landscape, reducing microwave absorption and thereby enhancing the Q-factor.

Interestingly, electron spin resonance (ESR) spectroscopy of this crystal revealed no paramagnetic impurities at 15 GHz. The closest zero-field splitting is observed at 11.31 GHz, which is most likely due to chromium ion impurities. This suggests that the observed microwave effects are not driven by paramagnetic defects but rather by an intrinsic dielectric response influenced by UV illumination.

The ability to control microwave modes with low-intensity light opens exciting possibilities for sensing applications, quantum technologies, and adaptive photonic systems. Rutile, a low-loss dielectric with a high permittivity (~100 at microwave frequencies), is particularly attractive for the miniaturization of microwave devices, as it allows for compact resonators with enhanced field confinement. Additionally, rutile is already widely used in UV-related applications due to its strong absorption and photocatalytic properties, making it a promising material for optically tunable microwave components. By leveraging light-induced modifications in its microwave properties, it becomes possible to create optically reconfigurable resonators, enabling precision sensing of environmental changes such as temperature, pressure, or charge carrier dynamics. This could further benefit quantum information processing, where controlled interactions between optical and microwave fields are crucial. Moreover, in metrology and spectroscopy, optically tunable rutile-based microwave cavities could enhance timekeeping, material characterization, and fundamental physics experiments, providing a dynamic, contact-free method for fine-tuning system properties. The unique combination of high permittivity, low loss, and UV sensitivity makes rutile an exceptionally promising material for next-generation tunable microwave technologies.

\section*{Acknowledgments}

This work was funded by the Australian Research Council Centre of Excellence for Engineered Quantum Systems, CE170100009 and Centre of Excellence for Dark Matter Particle Physics, CE200100008.

\vspace{10mm}

\bibliographystyle{unsrt}

\end{document}